\documentclass[reprint,aps,prl,twocolumn,preprintnumbers,amsmath,amssymb,floatfix]{revtex4}

\usepackage{epsfig}
\usepackage[colorlinks=true,linkcolor=blue]{hyperref} 
\usepackage{xcolor}
\usepackage{ulem}
\usepackage{dcolumn}
\DeclareGraphicsExtensions{.pdf}
\usepackage{graphicx}
\usepackage{multirow}

\allowdisplaybreaks[3]

\begin{document}

\title{Determination of Fragmentation Functions from Charge Asymmetries in Hadron Production}

\author{Jun~Gao, ChongYang Liu, Bin Zhou}

\affiliation{
    State Key Laboratory of Dark Matter Physics, Shanghai Key Laboratory for Particle Physics and Cosmology, Key Laboratory for Particle Astrophysics and Cosmology (MOE), School of Physics and Astronomy, Shanghai Jiao Tong University, Shanghai 200240, China
 }

\email{{jung49@sjtu.edu.cn}\\
{liucy1999@sjtu.edu.cn}\\
{zb0429@sjtu.edu.cn}\\
}

\begin{abstract}
We propose a novel method for extracting non-singlet (NS) fragmentation functions (FFs) of light charged hadrons from charge asymmetries measured in hadron fragmentation, using data from both single-inclusive electron-positron annihilation and semi-inclusive deep-inelastic scattering processes.
We determine the NS FFs for pions and kaons at next-to-next-to-leading order in Quantum Chromodynamics, including a comprehensive uncertainty analysis.
The extracted FFs reveal a scaling index of about 0.7 at large momentum fractions and low energy scales, a strangeness suppression factor of about 0.5, and universality in fragmentation of light mesons.
Our findings provide a valuable benchmark for testing non-perturbative QCD models and Monte Carlo event generators, and serve as crucial input for future electron-ion colliders.  
\end{abstract}
 \maketitle

\pagebreak
\newpage

\noindent{\it Introduction.--}
Fragmentation functions (FFs) are fundamental quantities in high-energy physics that describe the probability density in the transition of a parton into a specific hadron, differential in the light-cone momentum fraction of the parton carried by the hadron.
The study of FFs has evolved from early parton models, such as the Field-Feynman model~\cite{Berman:1971xz,Field:1977fa,Feynman:1978dt}, to the modern framework of QCD collinear factorization~\cite{Collins:1989gx,Metz:2016swz}.  
FFs are essential non-perturbative inputs for investigating the internal structure of nucleons and correlations within nuclei. 
This includes their application in determining polarized parton distribution functions (PDFs)~\cite{Borsa:2024mss,Cruz-Martinez:2025ahf} and nuclear PDFs~\cite{AbdulKhalek:2020yuc,Muzakka:2022wey,Eskola:2021nhw,Helenius:2021tof}.  
The precise determination of FFs is particularly critical for the upcoming era of high-precision nuclear physics, driven by the development of electron-ion colliders (EICs)~\cite{Accardi:2012qut,Anderle:2021wcy}. 
Furthermore, FFs extracted from high-energy data offer a significant opportunity to understand hadronization and confinement.
Global fits that combine perturbative QCD calculations with diverse experimental measurements have made it possible to extract FFs, especially of light charged hadrons. 
Notable efforts in this area include DSS \cite{deFlorian:2007ekg}, HKNS \cite{Hirai:2007cx}, AKK \cite{Albino:2008fy}, NNFF \cite{Bertone:2018ecm}, MAPFF~\cite{Khalek:2021gxf}, JAM \cite{Moffat:2021dji}, and NPC23~\cite{Gao:2024nkz,Gao:2024dbv,Gao:2025bko}.
These analyses were performed at next-to-leading order (NLO) in QCD, utilizing different data sets and theoretical prescriptions.
Recently, significant progress has been made toward determining FFs at next-to-next-to-leading order (NNLO). 
These efforts have used data solely from single-inclusive electron-positron annihilation (SIA)~\cite{Bertone:2017tyb,Soleymaninia:2018uiv}, or combined SIA and semi-inclusive deep-inelastic scattering (SIDIS) data at approximate~\cite{Borsa:2022vvp,AbdulKhalek:2022laj} or full~\cite{Gao:2025hlm} NNLO.  
However, the relationship between FFs extracted from high-energy data and non-perturbative QCD models remains largely unexplored.
In this work, we propose a direct extraction of the non-singlet (NS) FFs, defined as the difference between quark and anti-quark FFs, from the charge asymmetry measured in pion and kaon production in SIA and SIDIS processes. 
The availability of world data, combined with state-of-the-art NNLO perturbative QCD calculations, enables a precise and robust determination of the NS FFs for charged pions and kaons. 
We perform a pion-only fit and demonstrate that a simple three-parameter model for the NS FFs accurately describes all SIDIS data.  
The scaling index $\beta$ of the FFs at large momentum fractions is extracted, with a thorough assessment of uncertainties. 
Our results support the prediction of $\beta\sim 1$ from Nambu-Jona-Lasinio (NJL) model~\cite{Shigetani:1993dx,Heinzl:2000ht,Ito:2009zc,Holt:2010vj} and disagree with the predictions of $\beta\sim 2$ based on either perturbative QCD~\cite{Gunion:1980wt,Brodsky:1994kg,Yuan:2003fs,Ji:2004hz} or Dyson-Schwinger equations~\cite{Holt:2010vj,Shi:2018mcb,Cui:2021mom}. 
In a joint fit of both pion and kaon, we determine a strangeness suppression factor of about 0.5 and observe consistency in the NS FFs for fragmentation into pions and kaons.
We note there exist previous determinations on non-singlet FFs of kaons from $u$ and $d$ quarks utilizing SIA data on charged and neutral kaons production with assumptions of isospin symmetry~\cite{Albino:2010jc,Christova:2008te,Christova:2006qs}.
In comparison, our work represents a direct determination of NS FFs with all available world data and less model-dependence.
\noindent{\it Theoretical setup and data characteristics.--}
We consider charged hadron production via fragmentation in neutral current (NC) or (anti-)neutrino charged current (CC) SIDIS on target nucleus $H$, and SIA at the $Z$ boson mass pole.
Based on QCD collinear factorization~\cite{Metz:2016swz}, theoretical predictions on charge asymmetry of hadron production cross sections can be expressed as convolutions of PDFs $f_{i/H}$, FFs $D^{h^+}_{i}$ and non-singlet coefficient functions $\mathcal A_{NS}$,  
\begin{align}
\sigma_{h^+}-\sigma_{h^-}|_{NC}&=\sum_{i=u,d,s}\Big(f_{i/H}(x)-f_{\bar i/H}(x)\Big)\nonumber \\
&\qquad \otimes D^{h^+}_{i^-}(z)\otimes \mathcal A_{NS}^{NC}(z,x,i),\nonumber \\
\sigma_{h^+}-\sigma_{h^-}|_{\nu CC}&=\sum_{i=d,s}\Big(f_{i/H}(x)\otimes D^{h^+}_{i'^-}(z)\otimes \mathcal A_{NS}^{\nu CC,q}(z,x)\nonumber \\
 &-f_{\bar {i'}/H}(x)\otimes D^{h^+}_{i^-}(z)\otimes \mathcal A_{NS}^{\nu CC,\bar q}(z,x)\Big),\nonumber\\
 \sigma_{h^+}-\sigma_{h^-}|_{SIA}&=\sum_{i=u,d,s}D^{h^+}_{i^-}(z)\otimes \Big(\mathcal A_{NS}^{SIA,q}(z,\cos\theta,i)\nonumber \\
 &-A_{NS}^{SIA,\bar q}(z,\cos\theta,i)\Big),
\end{align}
where $i$ and $h$ label parton flavors and hadron species, respectively.
We have assumed charge conjugation symmetry for the fragmentation functions, i.e., $D^{h^+}_{i}(z)=D^{h^-}_{\bar i}(z)$, and have suppressed the dependence on QCD scales in the formulas above. 
The kinematic variables $x$, $z$ and $\theta$ are Bjorken variable, hadron energy fraction, and hadron polar angle, respectively.
The non-singlet fragmentation functions are defined as
$D^{h^+}_{i^-}(z)\equiv D^{h^+}_{i}(z)-D^{h^+}_{\bar i}(z)$.
In the case of CC SIDIS, $i'={u,c}$ for $i={d,s}$ when assuming diagonal CKM matrix. 
The non-singlet coefficient functions $\mathcal A_{NS}$ have been calculated to NNLO for SIDIS~\cite{Goyal:2023zdi,Bonino:2024qbh,Goyal:2024emo,Bonino:2025tnf,Bonino:2025qta} and SIA~\cite{Rijken:1996vr,Rijken:1996npa,Rijken:1996ns,Mitov:2006wy,Soar:2009yh,Almasy:2011eq,Xu:2024rbt,He:2025hin}.
\begin{table}[h]
\begin{tabular}{|c|c|c|c|c|}
\hline
  $zD^{\pi+}_{u^-}$      &  $\alpha$   & $\beta$   & $a_0(a_1)$      & $\chi^2$ ($N_{pt}$)  \\
   \hline
nominal       & $-0.335^{+0.22}_{-0.21}$  & $0.692^{+0.20}_{-0.19}$  & $-1.891^{+0.30}_{-0.28}$        & 252.4 (249)             \\ \hline 
mod.1           & \textbf{0}       & 0.976  & -1.438        & 257.5 (249)             \\ 
mod.2           & -0.035 & \textbf{1.0}    & -1.460        & 257.4 (249)             \\ 
mod.3           & 0.893   & \textbf{2.0}    & -0.091        & 326.4 (249)             \\ 
mod.4           & \textbf{1.0}     & \textbf{2.0}    & \textbf{0.054}        & 346.5 (249)             \\  
mod.5           & \textbf{0.681} & \textbf{2.298} & \textbf{-1.352(2.444)} & 563.2 (249)             \\  \hline
sys.1     & -0.323  & 0.703  & -1.846     & 247.6 (249)             \\ 
sys.2           & -0.476  & 0.665  & -1.984       & 152.7 (144)             \\ 
sys.3          & -0.355  & 0.732  & -1.828        & 153.8 (136)             \\ 
sys.4           & -1.000  & 0.835  & -3.919(2.350)   & 252.2 (249)             \\ 
sys.5           & -0.349  & 0.568  & -1.885       & 252.4 (249)             \\ 
sys.6           & -0.373  & 0.534  & -1.954      & 243.3 (249)             \\ \hline
\end{tabular}
  \caption{
  Fit quality for various models and systematic checks.
  Parameters shown in bold are fixed, not fitted.
  }\label{tab:chi2}
\end{table}
In this study we have analyzed a comprehensive set of data on identified $\pi^+$, $\pi^-$, $K^+$ and $K^-$ production from SIDIS and SIA processes.
The HERMES experiment measured pion and kaon production in NC SIDIS on both proton and deuteron targets at a center-of-mass energy of 7.26 GeV, with $z$ values up to 0.8~\cite{HERMES:2012uyd}.
The COMPASS experiment measured pion and kaon production in NC SIDIS on isoscalar targets~\cite{COMPASS:2016xvm,COMPASS:2016crr} and, more recently, proton~\cite{Alexeev:2024krc} targets at a center-of-mass energy of 17.33 GeV, with $z$ values up to 0.85.
The ABCMO experiment measured pion production in (anti-)neutrino CC SIDIS on proton targets at a center-of-mass energy of about 8.8 GeV, with $z$ values up to 1~\cite{Aachen-Bonn-CERN-Munich-Oxford:1982jrr}.
Both HERMES and COMPASS measurements are differential in the momentum transfer $Q$, and we include only data with $Q>2$ GeV to ensure the validity of our perturbative calculations.
The ABCMO data on hadron multiplicity are integrated over $Q>1$ GeV.
The SLD collaboration measured charged pion and kaon multiplicities in the light quark-jet hemisphere at the $Z$-pole, using longitudinally polarized electron beams~\cite{SLD:2003ogn,SLD:1998coh}.
We calculated corresponding theoretical predictions at NNLO in QCD~\cite{thsld} using the projected-to-Born method~\cite{Cacciari:2015jma}. 
The reported multiplicity data for $h^+$ and $h^-$ are converted to charge asymmetry data by taking their differences, accounting for correlated systematic uncertainties where known. 
This study marks the first time that COMPASS data from proton targets, SLD charge asymmetry data, and ABCMO neutrino SIDIS data have been included in a global analysis of FFs at NNLO. 
We focus on charged pion and kaon production at large-$z$ values ($z>0.3$).
A simple three-parameter functional form for the non-singlet FFs at the initial scale $Q_0$ is well-motivated, 
\begin{equation}
z D_{i^-}^{h} (z, Q_0)
=
z^{\alpha} (1-z)^{\beta}
\exp(a_0).
\end{equation}
In our nominal fits we set $Q_0$=1.3 GeV.
In alternative four-parameter form $a_0$ is replaced with $a_0+a_1\sqrt z$.
We conduct two independent analyses: one fitting pion FFs and another jointly fitting pion and kaon FFs.
We consistently assume isospin symmetry for pion FFs, $D^{\pi^+}_{u^-}=-D^{\pi^+}_{d^-}$ and set the non-singlet FFs for non-constituent quarks to zero at the initial scale ($D^{\pi^+}_{(s/c/b)^-}=0$ and $D^{K^+}_{(d/c/b)^-}=0$).  
In the joint fit, we further assume that all non-singlet FFs share the same $\alpha$ and $\beta$ parameters, allowing only the normalization to differ due to quark masses. 
This results in three free parameters for our nominal pion-only fit ($D^{\pi^+}_{u^-}$) and five for our nominal joint fit ($D^{\pi^+}_{u^-}$, $D^{K^+}_{u^-}$ and $D^{K^-}_{s^-}$).
The non-singlet FFs are evolved to higher scales using three-loop
time-like splitting kernels~\cite{Mitov:2006ic,Moch:2007tx,Almasy:2011eq,Chen:2020uvt}, implemented in a modified version of HOPPET~\cite{Salam:2008qg}, to ensure NNLO consistency.
Differential cross sections are calculated at NNLO in QCD using the FMNLO program~\cite{Liu:2023fsq,Zhou:2024cyk}.
Unless otherwise specified, we use the CT18 NNLO PDFs with $\alpha_S(M_Z)=0.118$~\cite{Hou:2019efy} for calculations involving initial hadrons. 
The renormalization, factorization, and fragmentation scales are set equal to the momentum transfer $Q$ for both SIA and SIDIS, with scale variations incorporated into the covariance matrix for the $\chi^2$ calculations~\cite{Gao:2025hlm}.

\noindent{\it Results on pion-only fit.--}
We first perform a fit to the non-singlet FFs of pions using data from HERMES, COMPASS and ABCMO, as the SLD charge asymmetry data for pions have excessively large uncertainties. 
The extracted parameters and $\chi^2$ values are presented in Tab.~\ref{tab:chi2}.
Our nominal fit yields a global $\chi^2$ of 252.4 for 249 data points, indicating an excellent description of the charge asymmetry data with NNLO calculations and a three-parameter non-perturbative FF.
Detailed comparisons of the theory and data are also available (see Appendix).
The uncertainties of FF are determined using the Hessian method with a tolerance of $\Delta\chi^2=2.3$.
This tolerance value is estimated based on the agreements of individual data sets in the fit, as detailed in Ref.~\cite{Gao:2025hlm}.
The scaling index at large-$z$, $\beta$, is determined to be 0.69 with an uncertainty of about 0.2. 
We also test predictions from various models, as listed in Tab.~\ref{tab:chi2}.
In the first model, the parameter $\alpha$ is fixed to 0, while the other two parameters are fitted.
This results in a best-fit $\chi^2$ that is 5 units higher than our nominal fit, with $\beta$ close to 1.
The resulting FFs and $\chi^2$ are similar if we fix $\beta=1$, as suggested by the NJL model~\cite{Holt:2010vj}.
Conversely, fixing $\beta=2$, as suggested by~\cite{Ji:2004hz,Shi:2018mcb,Cui:2021mom}, increases the $\chi^2$ by about 74 units.
We emphasize that it is possible these predictions on the scaling power are only valid in a region of $x$ very close to one, e.g., $z > 0.9$, which are not well probed by current world data.  
The preference for $\beta=1$ over $\beta=2$ aligns with observations from global analyses of pion PDFs~\cite{Novikov:2020snp,Kotz:2023pbu}, and could be altered by the inclusion of threshold resummation effects~\cite{Aicher:2010cb,Barry:2021osv,Cui:2021mom}.
In the final two models, the non-singlet pion FFs are fully fixed, based on the Field-Feynman model~\cite{Field:1977fa} and the CSM FFs~\cite{Xing:2025eip}.
Note that the CSM FF was refitted using our four-parameter form, and an additional constant term was added for the FF from Field-Feynman model. 
Neither of these models provides a good description of the charge asymmetry data. 

\begin{figure}[h]
    \centering
    \includegraphics[width=\linewidth]{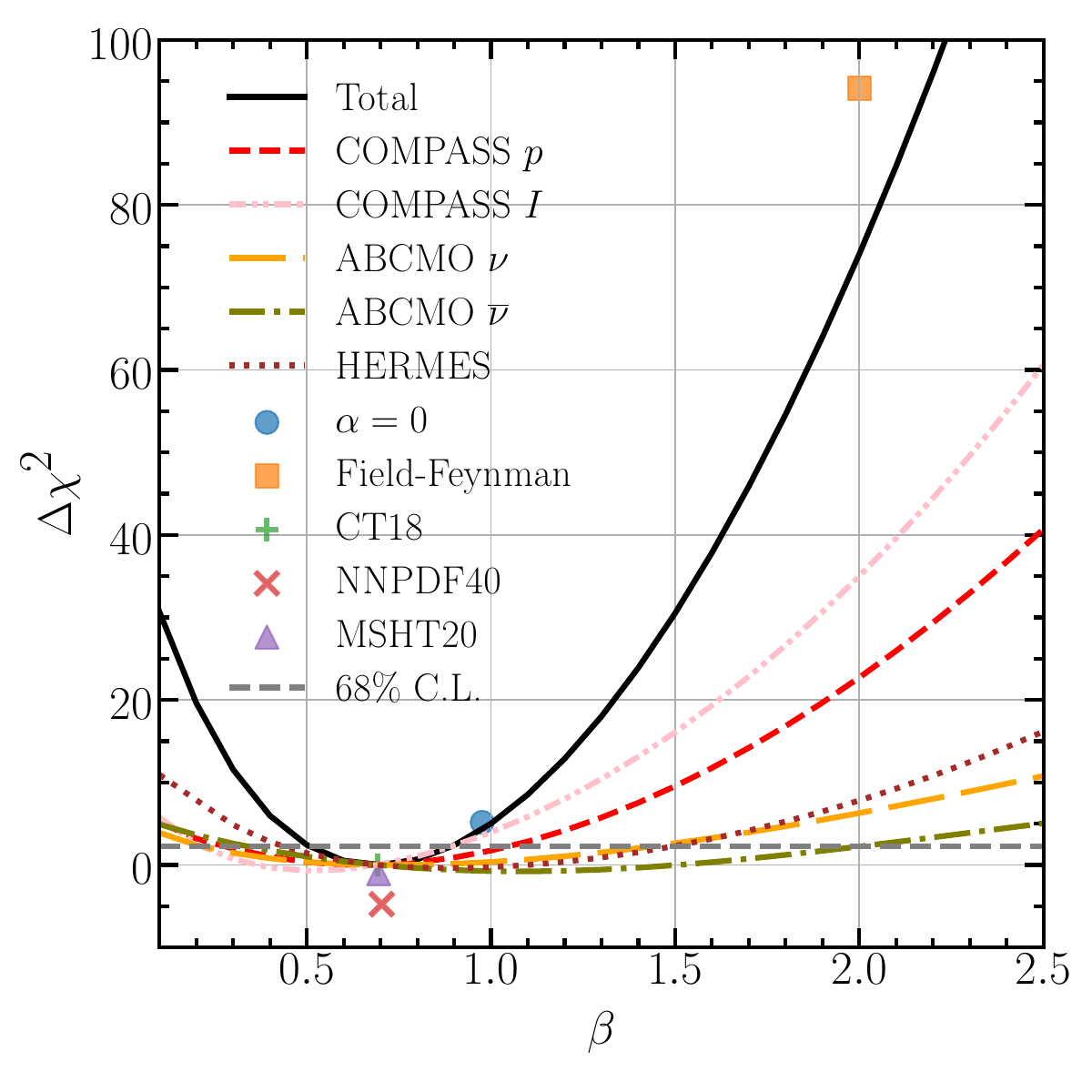}
     \caption{
     The global and individual $\chi^2$ variations as a function of the $\beta$ parameter. 
     The best-fit $\chi^2$ values from alternative fits and several models are also shown.
     }
    \label{fig:chi2}
\end{figure}

Alternative fits under varying conditions are summarized in Tab.~\ref{tab:chi2} to investigate various systematic effects.
First, we use NNPDF4.0 NNLO PDFs~\cite{NNPDF:2021njg} instead.
This yields a slightly lower $\chi^2$, but the extracted parameters are fully consistent with our nominal fit using CT18 PDFs.
In the other two fits, we raise the lower cut on $z$ to 0.5 or include only data from proton targets.
In both cases, the scaling index $\beta$ only changes slightly. 
Furthermore, we attempt a four-parameter fit or lower the initial scale to $Q_0=$ 1 GeV. 
The $\chi^2$ remains almost unchanged, while $\beta$ increases or decreases by about 0.13.
In the final scenario, we perform a NLO fit, which favors a $\beta$ value smaller by 0.16.
The NLO fit shows a slightly smaller $\chi^2$ than our nominal NNLO fit because the scale variations, accounted for in the $\chi^2$ calculation, are larger at NLO.
Additionally, we perform a dedicated scan of $\beta$ and plot the change in the global $\chi^2$ relative to our nominal best-fit as a function of $\beta$ in Fig.~\ref{fig:chi2}.
The horizontal dashed line indicates the global $\chi^2$ change corresponding to the 68\% C.L. interval.
The various markers represent the Field-Feynman model, the best-fit with $\alpha$ fixed to 0, and best-fits with alternative NNLO PDFs, including MSHT20~\cite{Bailey:2020ooq} and NNPDF4.0.
In Fig.~\ref{fig:chi2} we also plot the $\chi^2$ changes for individual data sets as a function of $\beta$ during the scan.
All three data sets show consistency, with preferred $\beta$ values falling within the uncertainty range of the global determination.
The COMPASS measurements on proton and isoscalar targets, with a total of 209 data points, provide the strongest constraints, followed by the ABCMO (28 data points) and HERMES (12 data points) measurements.

\begin{figure}[h]
    \centering
    \includegraphics[width=\linewidth]{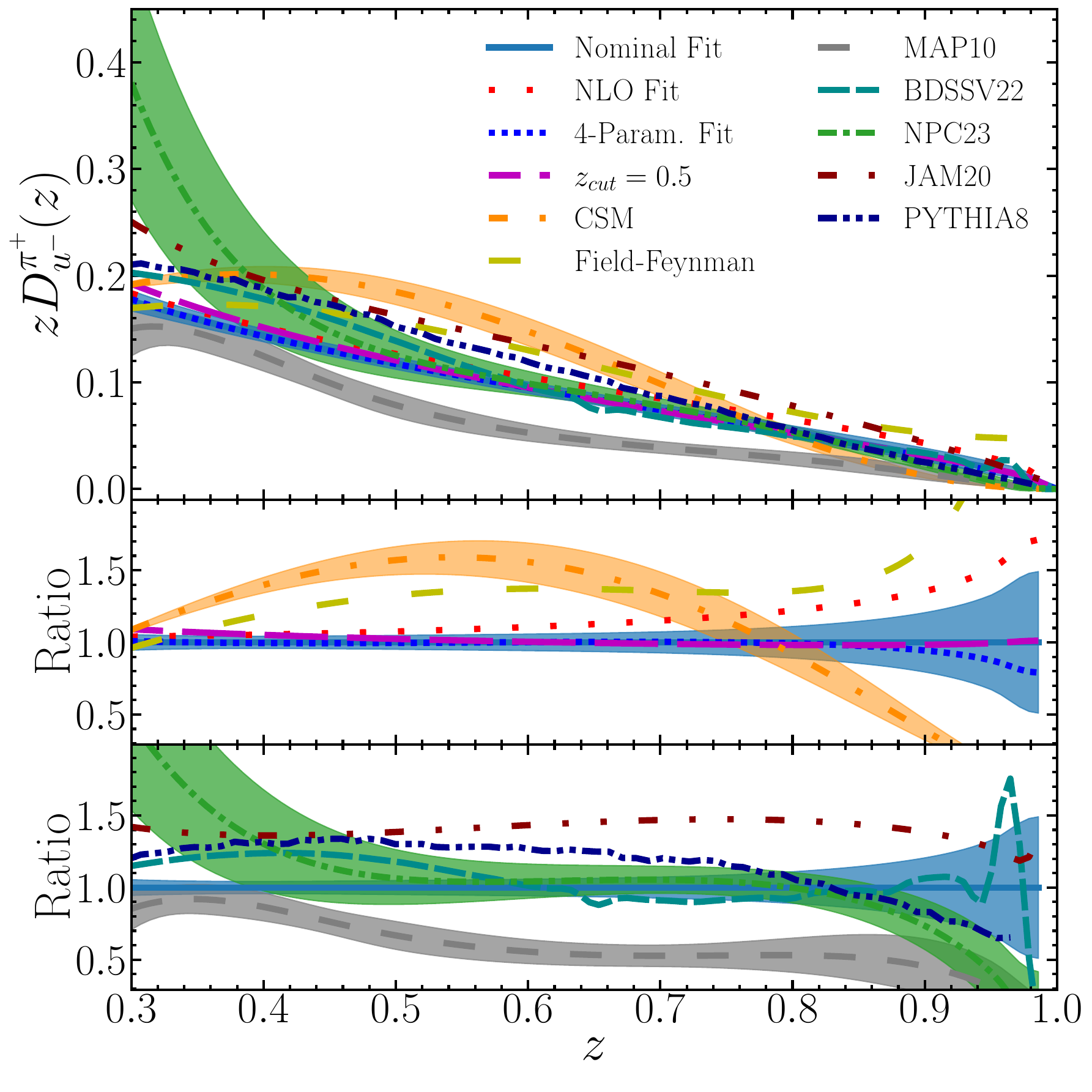}
     \caption{
     The NS FF $D_{u^-}^{\pi^+}$ at a scale of 1.3 GeV, as a function of $z$ from our nominal and alternative fits, compared with predictions from the Field-Feynman model, MC generators, and previous FF determination.
     }
    \label{fig:ffs}
\end{figure}

Our non-singlet pion FFs, $zD_{u^-}^{\pi^+}$, are shown in Fig.~\ref{fig:ffs} as a function of $z$ at a fragmentation scale of $1.3$ GeV.
This includes results from the nominal NNLO fit with Hessian uncertainties and from alternative fits under various conditions.
Our results are compared to previous determinations, including the Field-Feynman model, CSM FFs, FFs from the BDSSV22~\cite{Borsa:2022vvp}, MAP10~\cite{AbdulKhalek:2022laj}, and NPC23~\cite{Gao:2025hlm} global analyses at NNLO, and the JAM20~\cite{Moffat:2021dji} global analysis at NLO.
We also show predictions based on MC simulations with PYTHIA8~\cite{Bierlich:2022pfr}.
For the simulation, we select electron-positron collisions at a center of mass energy of 11 GeV; the results are insensitive to this choice.
Our NLO fit favors a slightly harder distribution compared to the NNLO fit. 
The CSM results include an uncertainty that reflects possible variation in the value of the hadron scale, and exhibit a very different shape with respect to our nominal results.
The FFs from the Field-Feynman model deviate significantly in the region $z>0.8$ due to the constant term not vanishing at $z=1$.
The BDSSV22 and NPC23 FFs are in good agreement with our nominal results, considering their respective uncertainties.
The PYTHIA8 predictions show differences in the shape of the FFs, and the MAP10~(JAM20) FFs are significantly below~(above) all other predictions.

\noindent{\it Results on joint fit.--}
In this section, we report the results of the joint fit to non-singlet FFs of pions and kaons using all aforementioned charge asymmetry data.
In this joint fit, the SLD kaon data provide direct and dominant constraints on the FFs from strange quarks, specifically $D_{s^-}^{K^-}$.
We have raised the $z$ cut to 0.4 in the nominal joint fit to account for the larger mass of kaons.
The best-fit yields a global $\chi^2$ of 436.0 for 378 data points with 5 free parameters.
The scaling index increases slightly to 0.79 for the joint fit.

\begin{figure}[h]
    \centering
    \includegraphics[width=\linewidth]{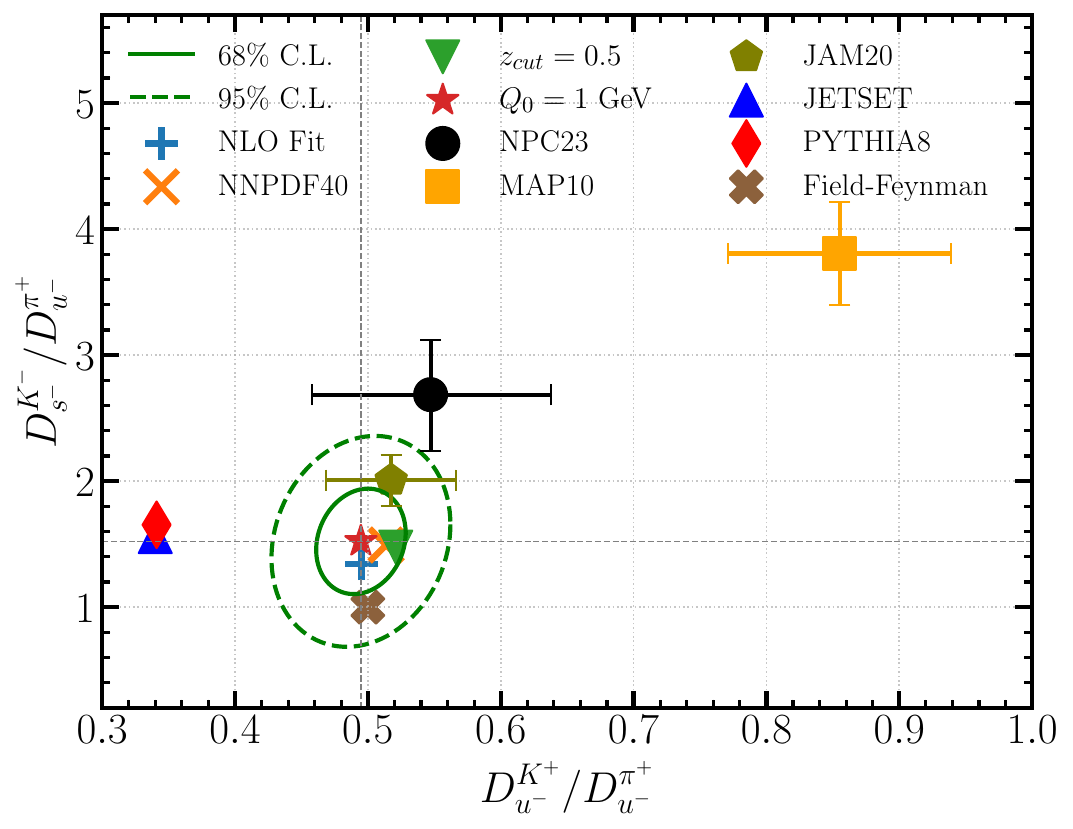}
     \caption{
     Ratios of the NS FFs of kaons and pions  at a scale of 1.3 GeV, from our nominal and alternative fits, compared with predictions from the Field-Feynman model, MC generators, and previous FF determinations.
     }
    \label{fig:ratio}
\end{figure}

We investigate the effects of quark mass on the FFs of light mesons, such as the strangeness suppression factor.   
In Fig.~\ref{fig:ratio}, we plot the ratios of the three NS FFs at 1.3 GeV: $D_{u^-}^{K^+}/D_{u^-}^{\pi^+}$ and $D_{s^-}^{K^-}/D_{u^-}^{\pi^+}$.
Our best-fit predicts these ratios to be 0.49 and 1.52, respectively. 
The two ellipses represent the 68\% and 95\% C.L. region determined with the Hessian method.
A tolerance of $\Delta \chi^2=5.4$ is estimated for our nominal joint fit due to the slightly worse agreement with the COMPASS proton data.
We show predictions from alternative fits with NNPDF4.0 PDFs, a $z$ cut of 0.5, and $Q_0=1$ GeV, and a NLO fit.
They all fall within the 68\% C.L. error ellipse of the NNLO nominal fit.
The Field-Feynman model predicts the two ratios to be 0.50 and 1.0, which is close to the boundary of the 68\% C.L. error ellipse.
We also calculate the ratios using either NNLO and NLO FFs from global analyses or MC simulations following similar procedure as explained in the pion-only fit.
In the calculations we have integrated the FFs over $z$ from 0.4 to 1 before taking the ratios. 
The central prediction from NPC23~(JAM20) FFs is close to the boundary of the 95\%~(68\%) C.L. error ellipse while that from MAP10 shows much larger values of the two ratios.
Both PYTHIA8 and JETSET~\cite{Sjostrand:1993yb} predict a stronger strangeness suppression with $D_{u^-}^{K^+}/D_{u^-}^{\pi^+}$ of about 0.34.

\begin{figure}[h]
    \centering
    \includegraphics[width=\linewidth]{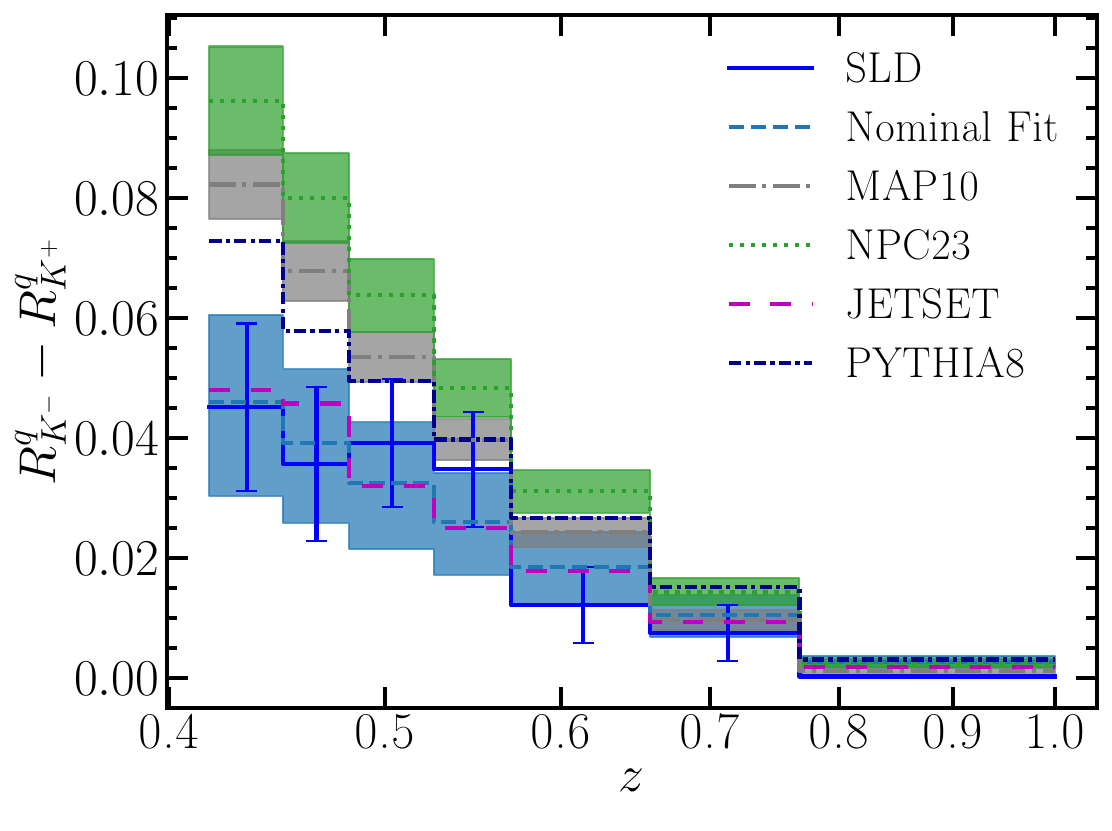}
    \caption{
    Comparison of SLD measurements of kaon charge asymmetry in the light-quark tagged hemisphere with theoretical predictions at NNLO based on FFs from this fit and from previous determinations, as well as predictions from MC event generators.
    }
    \label{fig:sld}
\end{figure}

In Fig.~\ref{fig:sld}, we highlight the impact of the SLD measurements of charge asymmetry in the light-quark tagged hemisphere, specifically $R^q_{K^-}-R^q_{K^+}$, where $R$ denotes the measured hadron multiplicities.  
We compare our NNLO predictions from the joint fit, including Hessian uncertainties, to the SLD data, which are shown with error bars.
They agree well within the large uncertainties on both sides.
Predictions from the JETSET simulation also agree well with the SLD data, due to its generally lower hadron multiplicities. 
Predictions from the PYTHIA8 simulation, as well as NNLO predictions using NPC23 and MAP10 FFs, are well above the SLD measurements.
\noindent{\it Conclusions.--}
We have presented a new approach to extracting the non-singlet fragmentation functions of charged pions and kaons at NNLO in QCD from charge asymmetries measured in both SIDIS and SIA processes.
This is the first time that charge asymmetry data from SIA and neutrino SIDIS have been included in a global analysis alongside theoretical calculations at NNLO accuracy. 
This methodology has led to a robust and precise determination of the NS FFs of light mesons, yielding several key findings. 
These include a large-$z$ scaling index close to 1, a strangeness suppression factor of approximately 0.5, and the universality of pion and kaon FFs. 
The extracted NS FFs were compared to those from previous global analyses and predictions from MC event generators, revealing notable differences.  
Our results serve as an important benchmark for testing non-perturbative QCD models and as crucial input for future electron-ion colliders. 

\quad \\
\noindent \textbf{Acknowledgments.} 
We thank M. Lochner and L. Bonino for their communications regarding the NNLO calculations of neutrino SIDIS, C. Roberts for providing numerical results of CSM FFs and comments on the draft, T. Sjostrand for communications on the JETSET predictions, and other members of the NPC collaboration for valuable discussions.
The work of JG is supported by the National Natural Science Foundation of China (NSFC) under Grant No.~12275173, the Shanghai Municipal Education Commission under Grant No.~2024AIZD007, and open fund of Key Laboratory of Atomic and Subatomic Structure and Quantum Control (Ministry of Education).

\section{Appendix: Comparison between Theory and Experimental Data}

\begin{figure}[hbtp]
    \centering
    \includegraphics[width=0.8\linewidth]{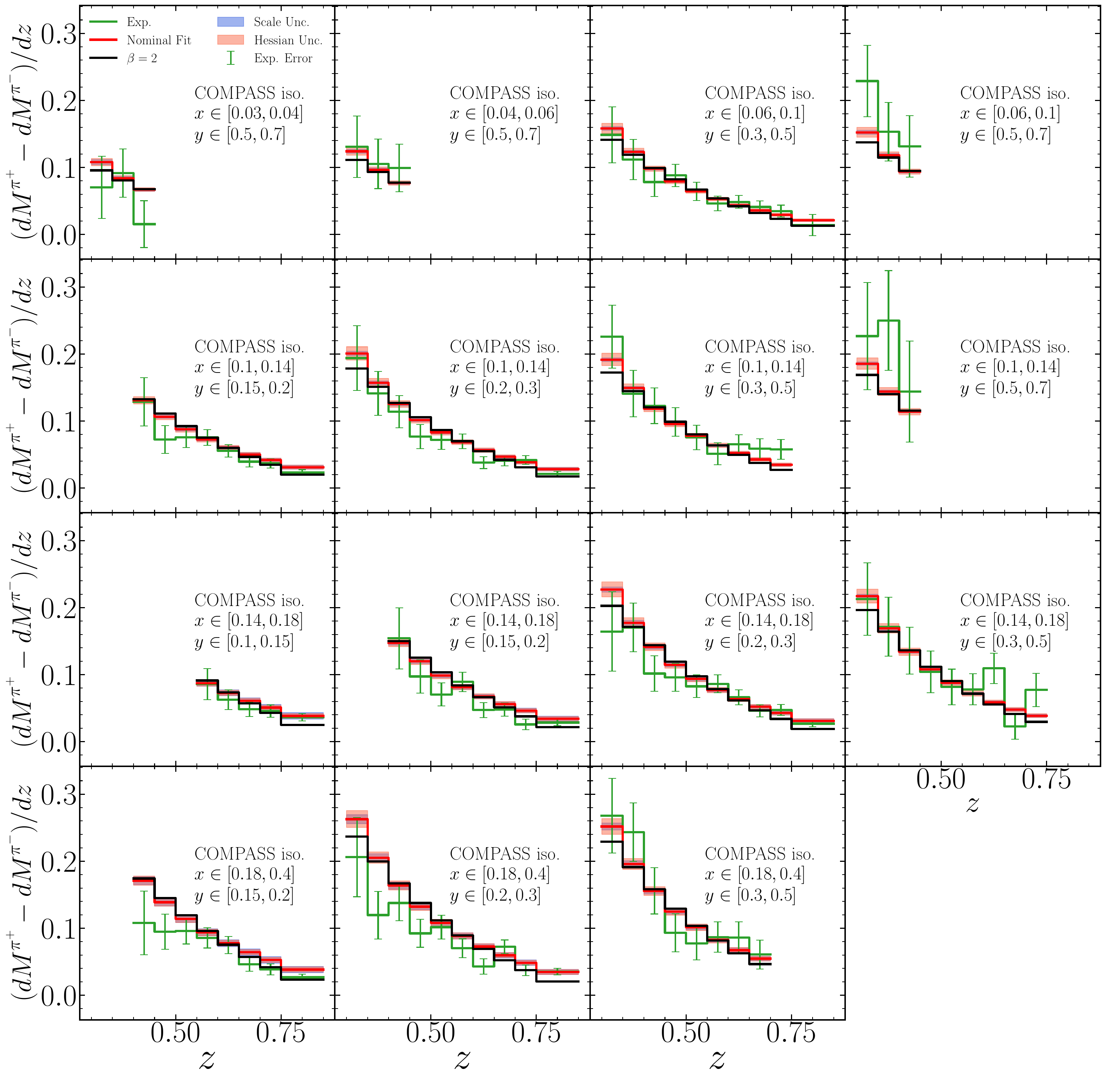}
    \caption{Comparisons of experimental data and predictions from nominal fit and fit fixing $\beta=2$ for COMPASS measurements of pion multiplicity difference with iso-scalar target. Scale uncertainties are shown in shaded bands and estimated by varying $\mu_R/{\mu_{R,0}}=\mu_F/{\mu_{F,0}}=\mu_D/\mu_{D,0}=\{1/2,1,2\}$ and taking the envelope. Hessian uncertainties are evaluated by the Hessian method. Experimental uncertainties shown are quadratic sum of statistical and uncorrelated systematic uncertainties. Each panel represents different kinematic regions in $x$ and $y$.}
    \label{fig:compassDT}
\end{figure}

\begin{figure}[hbtp]
    \centering
    \includegraphics[width=0.8\linewidth]{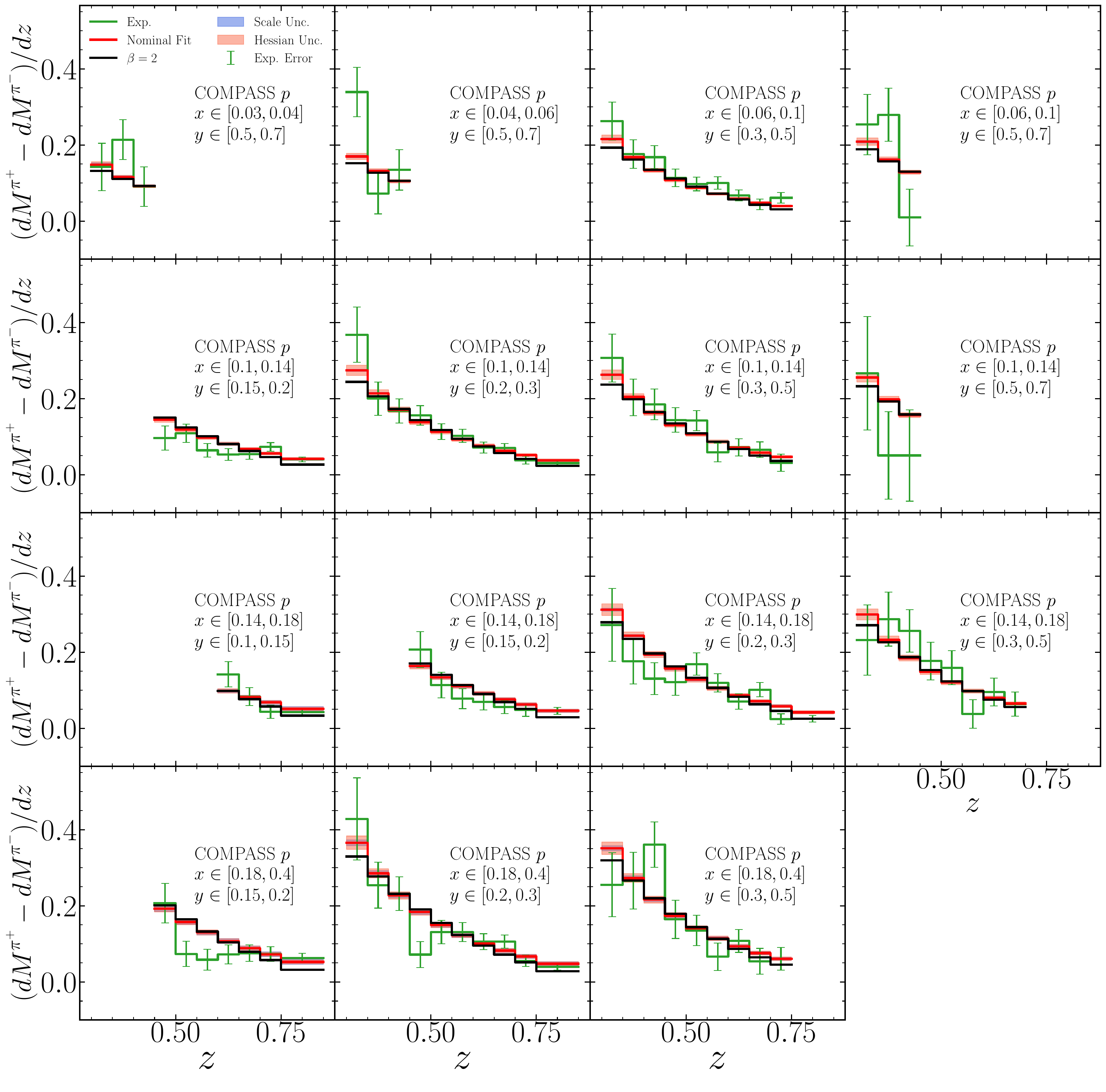}
    \caption{Same as Fig.~\ref{fig:compassDT} but for COMPASS measurements with proton target.}
    \label{fig:compassPT}
\end{figure}

\begin{figure}[hbtp]
    \centering
    \includegraphics[width=0.7\linewidth]{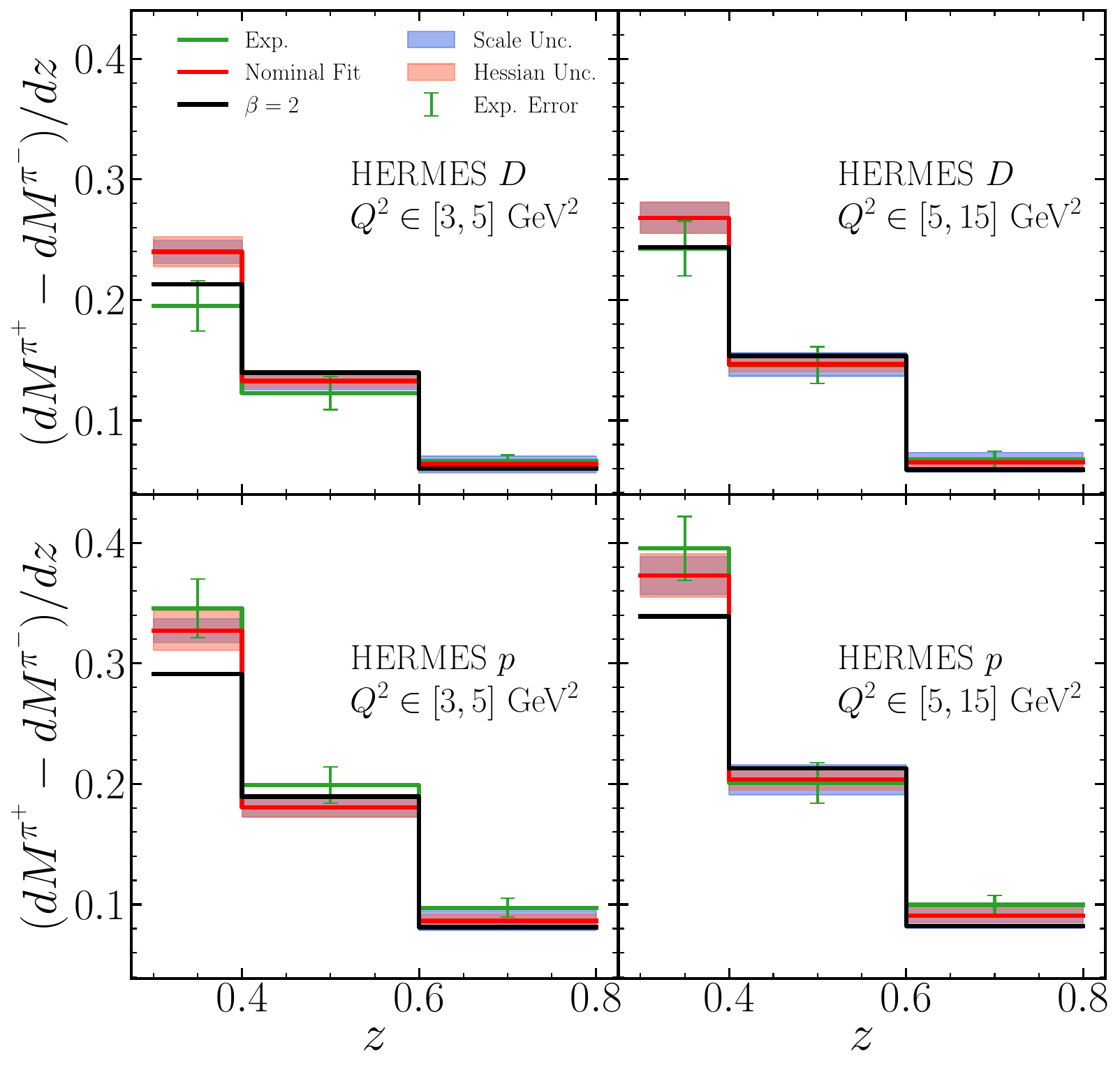}
    \caption{Same as Fig.~\ref{fig:compassDT} but for HERMES measurements with deuteron and proton targets respectively.}
    \label{fig:hermes}
\end{figure}

\begin{figure}[hbtp]
    \centering
    \includegraphics[width=0.7\linewidth]{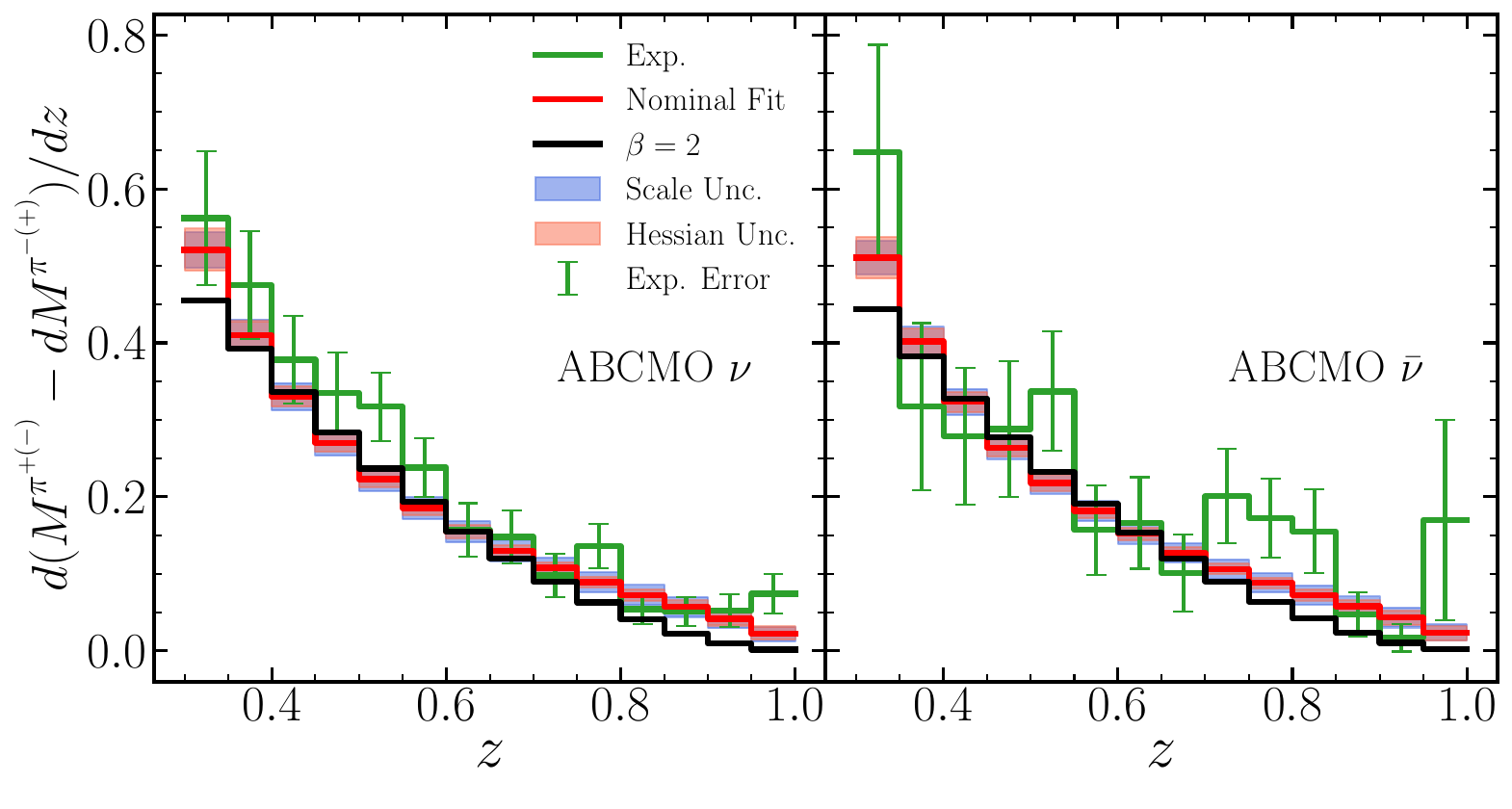}
    \caption{Same as Fig.~\ref{fig:compassDT} but for ABCMO measurements with $\nu$ and $\bar\nu$ beams.}
    \label{fig:abcmo}
\end{figure}

This analysis incorporates charged hadron multiplicity data from several key experiments. 
The COMPASS collaboration provides measurements on multiplicities of charged pions and kaons in kinematic bins of Bjorken-$x$ from $0.004$ to $0.4$ and inelasticity $y$ from $0.1$ to $0.7$.
This corresponds to a range in the squared momentum transfer, $Q^2$, from $0.36$ to $60~\text{GeV}^2$~\cite{COMPASS:2016xvm,COMPASS:2016crr}.
The measurements cover a region of the hadron energy fraction $z$ from $0.2$ to $0.85$ with both isoscalar and proton targets.
The HERMES collaboration has published charged pions and kaons multiplicities covering $x$ from $0.023$ to $0.6$, $z$ from $0.2$ to $0.8$, hadron transverse momentum $P_{h,\perp}$ from $0$ to $1.2~\text{GeV}$, and $Q^2$ from $1$ to $15~\text{GeV}^2$~\cite{HERMES:2012uyd}.
Additionally, we include charged pion production data from the ABCMO collaboration, which measured (anti-)neutrino charged-current semi-inclusive deep inelastic scattering on proton target~\cite{Aachen-Bonn-CERN-Munich-Oxford:1982jrr}.
These measurements were performed at a center-of-mass energy of approximately $8.8~\text{GeV}$, with kinematic cuts of $x>0.1$ and the hadronic invariant mass $W>3$ GeV, and $z$ values extending up to $1$.
In Figs.~\ref{fig:compassDT}-\ref{fig:abcmo} we present a comparison between the predictions of our nominal fit and the experimental data for each of the data sets used in our analysis. 
The experimental value and nominal fit prediction are shown in green and red lines respectively, while the theoretical uncertainty is shown by the shaded bands.
Scale variations are estimated by varying the renormalization ($\mu_R$), factorization ($\mu_F$), and fragmentation ($\mu_D$) scales simultaneously $\mu_R/\mu_{R,0}=\mu_F/\mu_{F,0}=\mu_D/\mu_{D,0}=\{1/2, 1, 2\}$ and taking the envelope.
Hessian uncertainties are determined using the Hessian method.
Experimental uncertainties shown are quadratic sum of statistical and uncorrelated systematic uncertainties.
We also display the predictions from an alternative fit fixing $\beta = 2$ as a reference, which results in worse description of the global data in general.
Figs.~\ref{fig:compassDT} and~\ref{fig:compassPT} show the comparison to COMPASS data from both isoscalar and proton targets across various kinematic bins.
Fig.~\ref{fig:hermes} presents the comparison with HERMES data from deuteron and proton targets in different $Q^2$ ranges.
Fig.~\ref{fig:abcmo} displays the comparison to ABCMO (anti-)neutrino-proton scattering data. For the $\bar{\nu}p$ scattering, the charged pion asymmetry is negative; we have plotted its absolute value for clarity.
Tabs.~\ref{tab:chi2_pi} and ~\ref{tab:chi2_pika} present the statistical measures of the fit quality. 
Tab.~\ref{tab:chi2_pi} corresponds to the pion-only fit, while Tab.~\ref{tab:chi2_pika} shows results for the joint pion and kaon fit. 
In the data sets of joint fit, the SLD 2004 measurement includes charged pion and kaon multiplicities in the light quark-jet hemisphere at the $Z$-pole, using longitudinally polarized electron beams~\cite{SLD:2003ogn}, while the 1999 measurement only includes charged kaon multiplicities~\cite{SLD:1998coh}.
We note that both of the SLD measurements presented unfolded results on hadron multiplicities corresponding to a pure light quark, according to the calculated purity of the light-quark jet.
The reported hadron multiplicities have been converted back to the level of a light-quark jet, in order to compare with our theoretical predictions at higher orders~\cite{thsld}.
For each data set, we report the number of data points $N_{pt}$, $\chi^2$, $\chi^2/N_{pt}$, and the effective Gaussian variable $S_E$. 
Note in estimation of the Hessian uncertainties, we use a $\Delta\chi^2$ tolerance being square of the maximal effective Gaussian variables of all individual data sets~\cite{Gao:2025hlm}.
For the pion-only fit, we also compare to the alternative fit fixing $\beta = 2$.

\begin{table*}[t]
  \begin{tabular}{|c|c||c|c|c|c|c|c|}
    \hline
      \multirow{2}{*}{pion ($z>0.3$)} &\multirow{2}{*}{$N_{\rm{pt}}$}  &  \multicolumn{3}{c|} {nominal fit} &  \multicolumn{3}{c|} {fixing $\beta=2$} \\
    \cline{3-8}
     & & $\chi^2$ &
    ${\chi^2}/{N_{\rm{pt}}}$ &
    $S_E$  & $\chi^2$ &
    ${\chi^2}/{N_{\rm{pt}}}$ &
    $S_E$ \\
    \hline
  HERMES ($p$)   & 6 & 2.5 & 0.42 &-1.11  & 12.6 & 2.10 & 1.64 \\
  HERMES ($D$)   & 6 & 6.7 & 1.12 & 0.38  & 4.4 & 0.74 & -0.30 \\
  COMPASS ($p$)  & 102 & 124.3 & 1.22 & 1.51  & 147.0 & 1.44 & 2.82 \\
  COMPASS (iso.) & 107 & 87.8 & 0.82 & -1.35  & 122.8 & 1.15 & 1.08 \\
  ABCMO ($\nu$)      & 14 & 15.3 & 1.09 & 0.37  & 21.6 & 1.54 & 1.36 \\
  ABCMO ($\bar \nu$) & 14 & 15.7 & 1.12 & 0.43  & 17.9 & 1.28 & 0.80 \\
    \hline
     \hline
     Total & 249 & 252.4 & 1.01 & 0.18  & 326.4 & 1.31 & 3.19 \\
    \hline
  \end{tabular}
  \caption{Quality of the pion-only fits. Results are shown for both the nominal fit and the constrained fit with $\beta = 2$. The number of data points $N_{pt}$, $\chi^2$, $\chi^2/N_{pt}$, and the effective Gaussian variable $S_E$ are listed for individual data sets.
}\label{tab:chi2_pi}
\end{table*}

\begin{table*}[t]
  \begin{tabular}{|c|c||c|c|c|}
    \hline
      \multirow{2}{*}{pion and kaon ($z>0.4$)} &\multirow{2}{*}{$N_{\rm{pt}}$}  &  \multicolumn{3}{c|} {nominal fit} \\
    \cline{3-5}
     & & $\chi^2$ &
    ${\chi^2}/{N_{\rm{pt}}}$ &
    $S_E$  \\
    \hline
  HERMES ($p$)   & 8 & 10.7 & 1.33 & 0.77  \\
  HERMES ($D$)   & 8 & 5.7 & 0.71 &-0.47 \\
  COMPASS ($p$)  & 153 & 196.4 & 1.28 & 2.32  \\
  COMPASS (iso.) & 169 & 185.0 & 1.09 & 0.88  \\
  ABCMO ($\nu$)      & 12 & 15.1 & 1.26 & 0.72  \\
  ABCMO ($\bar \nu$) & 12 & 13.2 & 1.10 & 0.37  \\
  SLD 1999 & 2 & 2.0 & 1.02 & 0.35  \\
  SLD 2004 & 14 & 7.9 & 0.56 &-1.25  \\
    \hline
     \hline
     Total & 378 & 436.0 & 1.15 & 2.03 \\
    \hline
  \end{tabular}
  \caption{Same as Tab.~\ref{tab:chi2_pi} but for pion and kaon joint fit. Only nominal fit is shown.}\label{tab:chi2_pika}
\end{table*}

\bibliography{main}

\end{document}